\documentclass[layout = twocolumn]{achemso}
\usepackage[utf8]{inputenc}
\usepackage[T1]{fontenc}

\usepackage{amsmath}
\usepackage{braket}
\usepackage{siunitx}

\usepackage{newtxtext,newtxmath}

\usepackage{microtype}
\usepackage{xcolor}

\newcommand{\E}{\mathrm{e}}
\newcommand{\im}{\mathrm{i}}
\newcommand{\vb}{\vec}
\DeclareMathOperator{\dif}{d}

\DeclareMathOperator{\tr}{tr}

\author{Edoardo G. Carnio}
\email{edoardo.carnio@physik.uni-freiburg.de}
\affiliation{Physikalisches Institut, Albert-Ludwigs-Universität Freiburg, Hermann-Herder-Str. 3,\\ 79104 Freiburg im Breisgau, Federal Republic of Germany}
\author{Heinz-Peter Breuer}
\affiliation{Physikalisches Institut, Albert-Ludwigs-Universität Freiburg, Hermann-Herder-Str. 3,\\ 79104 Freiburg im Breisgau, Federal Republic of Germany}
\alsoaffiliation{Freiburg Institute for Advanced Studies (FRIAS), Albert-Ludwigs-Universität Freiburg, Albertstr. 19, 79104 Freiburg im Breisgau, Federal Republic of Germany}
\author{Andreas Buchleitner}
\affiliation{Physikalisches Institut, Albert-Ludwigs-Universität Freiburg, Hermann-Herder-Str. 3,\\ 79104 Freiburg im Breisgau, Federal Republic of Germany}
\alsoaffiliation{Freiburg Institute for Advanced Studies (FRIAS), Albert-Ludwigs-Universität Freiburg, Albertstr. 19, 79104 Freiburg im Breisgau, Federal Republic of Germany}
\alsoaffiliation{Erwin Schrödinger International Institute for Mathematics and Physics, University of Vienna, Boltzmanngasse 9, 1090 Vienna, Austria}

\title{Wave-particle Duality in Complex Quantum Systems}

\begin{document}
\maketitle

\begin{abstract}
Stunning progresses in the experimental resolution and control of natural or man-made complex systems at the level of their quantum mechanical constituents raises the question, across diverse subdisciplines of physics, chemistry and biology, whether that fundamental quantum nature may condition the dynamical and functional system properties on mesoscopic if not macroscopic scales. But which are the distinctive signatures of quantum properties in complex systems, notably when modulated by environmental stochasticity and dynamical instabilities? It appears that, to settle this question across the above communities, a shared understanding is needed of the central feature of quantum mechanics: wave-particle duality. In this Perspective, we elaborate how randomness induced by this very quantum property can be discerned from the stochasticity ubiquitous in complex systems already on the classical level. We argue that in the study of increasingly complex systems such distinction requires the analysis of single incidents of quantum dynamical processes.
\end{abstract}

\begin{tocentry}
\includegraphics{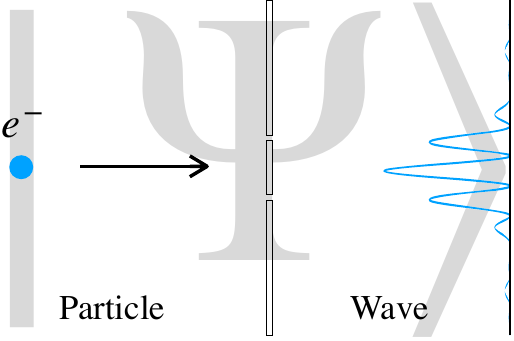}
\end{tocentry}

Which is the distinctive feature defining the demarcation line between the quantum and the classical world? When you raise this question in front of an even specialized audience of
mathematicians, physicists or chemists, you are likely to collect almost as many answers as people in the audience. Typical suggestions are ``interference'', ``superposition 
principle'', ``coherence'', ``entanglement'', and students in attendance often make the remarkable (at least for them) observation that the ``experts'' who teach their quantum mechanics 
classes apparently do not agree, or at least do not use a uniform language when addressing the very fundamental characteristics of one of the arguably most successful scientific theories
to date.
Given the constant specialization of scientific sub-disciplines, with mutually ever more hermetic jargons and little cross-talk (except through the above-mentioned students), this
was not perceived as a serious problem, as long as only the discourse within these sub-disciplines remained sufficiently consistent. However, with the last decades' improved 
experimental resolution and control of either natural or man-made complex systems with quantum constituents, from energy \cite{krueger2016}, charge \cite{deSio2018} 
and signal transport \cite{hore2017} in biological or condensed
matter, over radiation transport across multiply scattering materials \cite{labeyrie1999} to highly controlled, multi-constituent, cold quantum systems 
(with the quantum computer as just one 
potential application) \cite{Haroche2013a,Wineland2013a}, different communities focus, 
from different perspectives (and with different backgrounds), on the question of how non-trivial bona fide quantum effects can persist
in, and possibly determine the functional properties of, dynamically evolving, complex\footnote{``Complexity'' will here be tentatively 
understood as the manifest interplay of different energy and
time scales, preventing the identification of a preferred coordinate system (or basis set) for the deterministic description of distinctive system properties.} quantum systems\cite{Scholes2017}. 

Motivated by fruitful and educating, yet often controversial debates with colleagues from physical chemistry and biophysics, the present authors, originally trained in 
quantum optics, open quantum systems, quantum chaos and quantum transport theory, believe this to be the proper moment to attempt a concise description of our
current understanding of the distinctive features of quantum systems, and in particular of complex ones. Since our short answer to the initial question above is ``wave-particle
duality'', we hereafter attempt an explanation of what this means, in ever more complex settings, with a specific focus on the observable quantum properties of {\em single} quantum 
objects (``particles''!) or assemblies. Our presentation intentionally remains on an as elementary as possible level, to avoid fundamental principles to be overshadowed by technical overhead.

When dealing with atoms or molecules, one often thinks of a discrete spectrum of quantized energies $\varepsilon_i$ that an electron can access, and of the associated (orthogonal, hence mutually exclusive) orbitals $\chi_i$, themselves quantum states, which determine the probability amplitude to find the electron in a given region of space. 
An arbitrary electronic quantum state $\phi$ can then be represented by a linear combination of atomic orbitals \cite{Levine2009} (LCAO), $\phi = c_1 \chi_1 + c_2 \chi_2 + \ldots$, where the squared amplitudes $|c_i|^2$ of the complex coefficients $ c_i $ give the probability to detect the energy $ \varepsilon_i $. In more formal terms, to describe a quantum system's state we need a complete set of commuting \emph{observables} (CSCO -- as many are needed as the number of system degrees of freedom),
i.e.\ an exhaustive set of measurable system properties. So far, 
under the implicit assumption of a one degree of freedom system, our observable of choice 
has been the total energy (the Hamiltonian function), with $\varepsilon_i$ and $\chi_i$ its eigenvalues and eigenstates, respectively. Nothing prohibits us, of course, from 
employing a different CSCO, tantamount of choosing a different basis or coordinate system.

But what do we precisely mean when we talk of a ``quantum system'', and of the ``state'' it resides in? Quantum mechanics is often described as an intrinsically 
statistical theory, with states representing a statistical ensemble of faithful representatives of the given quantum system under study. 
Most prominently this was expressed by Schr\"odinger's 
affirmative statement that ``we are not \emph{experimenting} with single particles, any more than we can raise Ichthyosauria in the zoo.''\cite{Schrodinger1952}
Yet, Schr\"odinger's famous equation is perfectly deterministic, only the outcome of measurements -- even on single quantum objects in splendid isolation -- is not. 
That the predictions of quantum mechanics have to be taken seriously on the level of its single, elementary constituents like atoms, ions or photons has become a
solidly established fact only during the past, say, 30 years -- since the experimental monitoring of the quantum jumps of single ions \cite{Nagourney1986,Bergquist1986,Sauter1986}, of quantized resonator fields 
\cite{benson1994}, 
or of the Methuselah photon \cite{Gleyzes2007}.

Note that these experiments convey a two-fold message: First, on the fundamental level, they incarnate 
wave-particle duality through the statistics
of measurement results on {\em single} quantum objects,
being described by a linear evolution equation which exhibits the superposition principle
proper to any linear vector space structure -- here of the vector space of square integrable quantum states.
Second, it was this very demonstration of wave-particle duality on the most elementary level which {\em alone} allowed to 
even contemplate quantum computation devices as one of the potentially most revolutionary applications of modern 
quantum science.

Quantum computers, much as large molecular structures with strongly coupled sub-units and constituents, however, are systems with {\em many} degrees of freedom, by 
construction. The reason that the above wave-particle paradigm, so nicely and convincingly demonstrated for the elementary constituents of matter \cite{Merli1976}, still awaits
a truly convincing technological application to please a widespread utilitarian view of science, resides in the fact that the larger the number of system degrees of freedom (at least 
one per elementary constituent), the more difficult an accurate state preparation (due to the rapid growth of state space with the number of 
interconnected constituents) \cite{haeffner2005,brugger2018MSc}, and the more likely the impact of some residual coupling to external, 
uncontrolled, {\em environmental} degrees of 
freedom, since shielding a growing number of system degrees of freedom against the environment requires ever increasing experimental overhead.\footnote{This becomes very tangible in current efforts to run even small to moderate numbers of ``qubits'' coherently \cite{mielenz2016,abdelrahman2017} . Note that ``perfect'' shielding against environmental degrees of freedom always means control on {\em finite} time scales, and can, {\em in principle}, never be achieved forever, not even for a single degree of freedom. Hence the requirement to reach the ``strong coupling'' regime \cite{haroche1992}, to harvest on quantum resources.}

The physical term for the 
absence of accurate state control is ``disorder'', which is quantified by entropy 
measures of {\em classical} statistical randomness. Given a quantum system where such classical {\em randomness} co-exists with quantum mechanical {\em uncertainty},
an LCAO-type, vector-valued representation of the system state needs to be substituted by a {\em statistical} or {\em density operator} \cite{Neumann}, which indeed can account for both sources
of indeterminism on the measurement level. 
The system is then said to be in a {\em mixed state}, while any
quantum state which can be represented by a state vector is called {\em pure}. 

Note that there are two ways for mixed states to emerge in  
quantum mechanics: the first is the above type of classical randomness 
on the level of the system degrees of 
freedom only. Imagine a quantum mechanical protocol of state preparation either by measurement of a 
(non-degenerate) observable, or by the application of a multi-particle unitary with a finite set of control 
parameters. If, in the former case, the measurement record is deleted (what defines a {\em non-selective 
measurement}), or, in the latter, the control parameters cannot be specified with a certain critical level of precision, 
the 
prepared system state 
must be described by a set of possible state vectors occurring with certain probabilities.

A second way to realize a mixed state is to couple the quantum system to some 
external, equally quantized
degrees of freedom.
Thus, the system is not closed, but regarded as a subsystem of some larger system, 
making it an {\em open quantum system}. The structure of the Hilbert space comprising
external and system degrees of freedom immediately implies that, in general, there is no unique
mapping of a given pure state of external {\em and} system degrees of freedom onto a pure system state, 
but only on a mixed system state which is obtained by averaging over the populated states of the external degrees
of freedom (tantamount of taking the total state's partial trace over the latter). This structural feature is 
known as {\em entanglement} and {\em generically} induces statistical randomness on the system level while the state 
of system {\em and} external degrees of freedom is perfectly pure. Note that this randomness on the single-degree-of-freedom level is a consequence 
of the superposition principle, and, hence, of quantum mechanical uncertainty, in the case of systems with many degrees of freedom: 
Only when ignoring the state of the external degree(s)
of freedom (which may be unavoidable, if there is an overwhelming number of them) do we need 
to average over possible measurement outcomes on them.\footnote{This is the deep {\em physical} meaning of entanglement,
inseparable from the uncertain outcome of measurements on single degrees of freedom, and in this sense a manifestation 
of wave-particle duality on the many degrees of freedom level. Much more than just the formal non-separability of states on tensor product
spaces!} There are simply many more possible states 
which can be prepared by superposition of system and external degree of freedom states than just the products
of possible system and possible external states. Therefore, knowing everything that can be known about the state 
of system {\em and} external degrees of freedom forces us to admit a statistical distribution of compatible states
once we restrict our description to either sub-unit.

But not only does an increasing number of degrees of freedom enhance the system's  
sensitivity to disorder and/or environmental stochasticity,
thus enforcing its description by a density matrix, it generically also reduces the system's dynamical stability with respect to such perturbations, since many strongly coupled system degrees of freedom generically induce 
non-integrable\footnote{Classically, in the sense of Hamiltonian 
chaos \cite{percival1991}, or quantum mechanically, in the sense of random matrix theory (RMT) \cite{bohigas1991} or of modern semiclassics 
\cite{gutzwiller1991}.} 
spectral structures and, consequently,
dynamics. These are another source of -- deterministic and large -- fluctuations which may very strongly affect the counting statistics upon measurement, even on time scales 
when the environment's stochastic character only manifests as a slow drift of the boundary conditions the system degrees of freedom are subject to.

Therefore, when striving to harvest wave-particle duality on the level of large, multi-component and, possibly, multi-scale, composite quantum systems, whether 
quantum computing devices or macromolecular functional units, {\em we need the ability to discriminate quantum mechanical uncertainty from environmentally induced 
stochasticity, in the presence of dynamical instabilities}. Let us therefore first recollect the essential theory of quantum coherence \cite{mandelwolf,glauber1963}.

Given a finite-dimensional quantum system with Hilbert space spanned by a discrete basis 
defined by the eigenstates 
$ \lbrace \ket{\chi_1}, \ldots, \ket{\chi_N} \rbrace $ of 
an observable $A$, with 
associated (nondegenerate) eigenvalues $a_i$,\footnote{We restrict ourselves to a finite-dimensional, non-degenerate setting here, which 
suffices for our present purpose. More general treatments will not affect our qualitative conclusions.}
the density operator $ \rho $ can be written as an $ N \times N$ matrix. 
Its {\em diagonal} elements 
$ \rho_{ii} = \braket{\chi_i | \rho | \chi_i} $ are called \emph{populations}, because, if we measure $A$ on a sufficiently large number 
of identically prepared copies of the system, a fraction $ \rho_{ii} $ of them will yield the outcome $a_i$, and thereupon must 
be described
by $ \ket{\chi_i} $ after the measurement process. This is the essence of the projection postulate of quantum mechanics, and mathematically formalizes the 
information update inferred from each measurement result, on a given state $\rho$. More anthropomorphically, though also less precisely and easily giving rise to misled 
extrapolations
on some underlying ``ontology'', this is referred to as the ``collapse of the wave function''. Note, however, that nothing is collapsing here, only the number of 
possible a priori assumptions on the state under scrutiny is reduced -- through the registering of a concrete measurement event.\cite{englert2013}

The statistical operator's {\em off-diagonal} elements $ \rho_{ij} = \braket{\chi_i | \rho | \chi_j} $, instead, are called \emph{coherences}, because, when 
$ \rho_{ij} \neq 0 $, we can observe interference in experiments 
which probe the amplitudes of states $ \ket{\chi_i} $ and $ \ket{\chi_j} $. Most intuitively, both concepts can be rationalized by the minimalistic example of a {\em two-level system}
which can be realized by a spin-one-half particle like an atom, or by a many-particle system with an emerging, effective two-level structure in the 
studied degree of freedom.

Here we refer to the historical example realized
with the Stern-Gerlach apparatus \cite{Gerlach1922}, where a beam of silver atoms travels through an inhomogeneous magnetic field and is then detected on a screen. Because the total angular momentum of the atoms is determined by the spin of its valence $ 5s $ electron, the beam is deflected up- or downwards, such that only two spots are observed on the detector screen. This demonstrated, in 1922, the quantization of total angular momentum.
Formally, a general state is written as $ \ket{\Psi} = c_1 \ket{\chi_1} + c_2 \ket{\chi_2} $, 
with $ \ket{\chi_{1,2}}$ 
the eigenstates of the total angular momentum $ \mathcal{S}_z $ of the atom along the $ z $ axis. In this basis, 
the density operator 
reads
\begin{equation}
\rho = \ket{\Psi}\bra{\Psi} = \begin{pmatrix}
|c_1|^2 & c_1 c_2^* \\
c_2 c_1^* & |c_2|^2
\end{pmatrix} \, .
\end{equation}
The meaning of the coherence $ \rho_{12} $ becomes apparent when 
a measurement is performed which correlates 
the states $ \ket{\chi_1} $ and $ \ket{\chi_2} $. This is realised
when we measure another direction of the total angular momentum, for instance $ \mathcal{S}_x $, 
with eigenstates 
$ \ket{\phi_\pm} = (\ket{\chi_1}\pm\ket{\chi_2})/\sqrt{2} $. The probability 
to detect the associated eigenvalues $ \pm \frac{1}{2}$ is given by
\begin{equation}\label{eq:coh-sup}
p = \braket{ \phi_\pm | \rho | \phi_\pm} =
|\braket{\phi_{\pm} | \Psi} |^2 =
\frac{1}{2} \pm \Re(c_1^* c_2) \, .
\end{equation}
Because the term $ \rho_{21} = c_1^* c_2$ here enhances or reduces the individual probabilities $ |c_{i}|^2 $ by up to $\pm 100\ \%$, it is an \emph{interference} term.

The above example explicitly shows that 
coherences quantify 
the ability to 
exhibit interference in measurements which probe superpositions of 
eigenstates of a given observable,
such as $\mathcal{S}_z$. This is the very reason to represent the density operator in that very basis. 
Consequently, 
the concept of coherence is \emph{inherently} 
{\em basis-dependent}. 
On the other hand, 
the ability to 
exhibit interference must be encoded in the state itself, it cannot disappear with a change of basis, 
much as the physics 
cannot change with the frame of reference. This leads to the following consideration: 
A change of basis in the space of density operators can be written as $ \rho' = U \rho U^\dagger $, where $ U $ is a \emph{unitary} 
operator, i.e.\ it satisfies $UU^\dagger = U^\dagger U = \mathbb{I}$. Under unitary transformations, 
any power $k\in \mathbb{N}$ of $\rho$ therefore has invariant trace, thanks to its cyclic property: $ \tr\rho'^k = \tr \rho^k $.
This suggests $\rho^k$ as 
a good candidate to quantify the coherence of a state. Since $k=1$ fixes 
the state's normalization, 
$\tr \rho = 1$, the \emph{purity} 
$\tr \rho^2$ comes natural:
If the state is pure, like in the example above, 
it is projector-valued, $\rho = \ket{\Psi} \bra{\Psi}$, and, in particular, 
$\tr \rho^2 = \tr \ket{\Psi} \bra{\Psi} = 1$, which is the upper bound for the purity of any state. On the other hand, due to the above invariance under unitary transformations, 
the 
\emph{maximally mixed} state $\rho^\star = \mathbb{I}/N$ 
can be written as a matrix with constant populations and vanishing coherences in \emph{any} basis. For this reason, the purity of this state, 
which is $\tr(\mathbb{I}/N^2) = 1/N$, is the lower bound for the purity of any state of dimension $N$. All in all, the purity of a state $\rho$ is bounded by these two limits, 
$ 1/N \leq \tr \rho^2 \leq 1$, 
and therefore
effectively quantifies the coherence imprinted into the quantum state \cite{Jakob2007}.

States with purity less than unity are called \emph{mixed} states, and, 
as already discussed above, cannot be represented by a single state vector $\ket{\Psi}$, by purity's very definition.
The associated density operator is 
a \emph{statistical mixture} \cite{Neumann} of different pure states $ \ket{\Psi_k}$, with associated  
event probabilities $f_k$:
\begin{equation}\label{eq:stat-mixture}
\rho = \sum_k f_k \ket{\Psi_k} \bra{\Psi_k} \, .
\end{equation}
This is the mathematical expression of the fact that 
only partial information is available (and attainable!) on the state under study: only with 
some probability $f_k$ 
the system is prepared in the state $ \ket{\Psi_k}$.
In other words, one deals with an ensemble of quantum systems prepared, 
with relative abundance  
$f_k$, in either one of the states $\ket{\Psi_k}$.

Due to the available information on the state being incomplete, the statistical mixture (\ref{eq:stat-mixture})
can actually comprise 
different sets of (normalised) system states.
In particular, the $ \ket{\Psi_k} $ 
neither need
to be orthogonal, nor 
need they 
form a basis. 
Consequently, there are in general
infinitely many different mixtures which 
define one and the same density matrix $\rho$ \cite{Hughston1993,schroedinger1936}. 
The physical interpretation is that all these different mixtures exhibit identical statistical properties with regard to
arbitrary measurements, and there is no physical observable, i.e.\ no measurement set-up, which 
discriminates between these mixtures. 
Nevertheless, given a density matrix describing some quantum state, this can be
diagonalized to give 
$ \rho = \sum_i p_i \ket{i}\bra{i}$, where
the eigenstates $ \ket{i}$ \emph{do} form an 
orthonormal basis.
For a mixed state we must have more than one non-vanishing $p_i$, all less than unity 
due to normalization.
This proves our earlier statement that an arbitrary state's purity is always smaller than one, $\tr \rho^2 = \sum_i p_i^2 < 1$.

\begin{figure*}[t] 
   \centering
   \includegraphics[width=\textwidth]{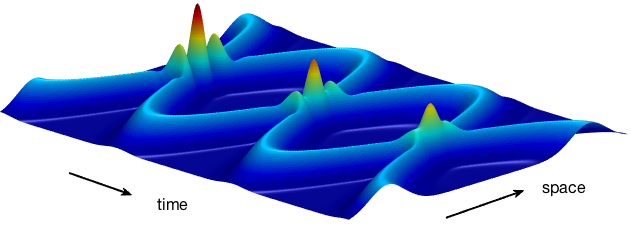} 
   \caption{Time evolution of a superposition of two wave packets 
in a harmonic potential. The picture shows the probability $p(x,t) =  \braket{x | \rho(t) | x}$ of finding the quantum system at position $x$, where the state of the system $\rho(t)$ is the solution of the master equation for a damped oscillator in a finite temperature 
reservoir \cite{breuerbook}. Whenever the superimposed wave packets merge, 
an interference pattern arises. Decoherence leads to a decrease 
of the interference contrast with increasing time.}
   \label{fig:interference}
\end{figure*}

All the physical processes that, in one way or 
another, reduce a state's purity and, hence, make it 
lose its ability to display interference, are usually called \emph{decoherence} phenomena (they induce a decay of the density operator's coherences in time, at least on average). An example is depicted in Fig. \ref{fig:interference}. The possible causes of decoherence are quite diverse, and we discuss some important examples in the following.

We start out with the case of  
a quantum system coupled to an environment\cite{breuerbook},
described, respectively, by $ \rho_\text{S} $ and $ \rho_\text{E} $, and we assume an initially uncorrelated state: $ \rho_\text{SE} = \rho_\text{S} \otimes \rho_\text{E} $. 
The composite state 
undergoes unitary time evolution, with a unitary operator $ U(t) $ acting on the 
entire Hilbert space $ \mathcal{H}_\text{SE} $, $ \rho_\text{SE}(t) = U(t) \rho_\text{SE} U^\dagger(t) $. 
When interested in the time evolution on the system degrees of freedom alone, the adequate mathematical object is the (reduced) system state 
$\rho_\text{S}(t)$, obtained upon tracing out the environmental degrees of freedom, which is tantamount to averaging over the possible states
of the environment. It turns out that the time evolution of $\rho_\text{S}(t)$ can be directly inferred from the system's initial state $\rho_\text{S}$,
with the help of the time-dependent 
\emph{Kraus operators} \cite{Kraus1983} $ E_k $,
as
\begin{equation}\label{dynmap}
\rho_\text{S}(t) = \tr_\text{E} \rho_\text{SE}(t) = \sum_k E_k(t) \rho_\text{S} E^\dagger_k(t) \; .
\end{equation}
We speak of decoherence when the coherences of $\rho_\text{S}(t)$ decay in time because of the 
non-unitary character of the dynamics (\ref{dynmap}), representing dissipative and irreversible processes
through the action of the Kraus operators $ E_k $.

Note that decoherence in this scenario requires non-vanishing 
coupling between 
system and 
environment. This coupling also introduces correlations within the total system state 
$\rho_\text{SE}$, which
may be a consequence of the entanglement of mutual degrees of freedom. However, entanglement between 
system and environment
is {\em not} a necessary prerequisite for decoherence. In fact, decoherence can also result from purely classical correlations, without 
any entanglement at any 
time. 

This can be illustrated by the simple example 
of a central spin model\cite{Schliemann2003}, 
where the principal system is given by one specific spin, defined by the operator $ \mathcal{S}_z $, and the environment is modeled by a bath of $ M $ further spins, 
each described by $ \mathcal{S}_z^{(k)} $ in their respective spaces. The total Hamiltonian then reads
\begin{equation}\label{eq:ham}
H_\text{SE} = \frac{\omega_0}{2} \mathcal{S}_z + \sum_{k=1}^{M} A_k \mathcal{S}_z \otimes \mathcal{S}_z^{(k)} \; ,
\end{equation}
where $ A_k $ are the coupling constants. 
If the central spin is initialized in the linear combination $ \ket{\Psi} = c_1 \ket{\chi_1} + c_2 \ket{\chi_2} $, while the environment is 
prepared
in the maximally mixed state $ \rho_\text{E}(0) = \mathbb{I}_\text{E}/2^M $,
$  H_\text{SE} $ induces a unitary evolution of system and environment which turns out to exhibit the structure
\begin{equation}
\rho_\text{SE}(t) = \frac{1}{2^M} \sum_{\vb{m} } \ket{\psi_{\vb{m}}(t) } \bra{\psi_{\vb{m}}(t)} \otimes \ket{\vb{m}} \bra{\vb{m}} \; .
\end{equation}
The vector $ \vb{m} $ describes the state of the environment spins, with components 
$ 0 $ (`spin down') 
or $ 1 $ (`spin up'). The time-evolved state of system and environment 
is therefore given by a statistical mixture of multipartite states $ \ket{\psi_{\vb{m}}(t) } \otimes \ket{\vb{m} } $, where 
{\em specific} central spin states 
are ``flagged'' by specific states of the environment. The environment 
acts as a probe which allows to perform an indirect measurement on the central spin, since 
the state of the environment immediately determines 
the system state. 
Because of the coupling in the Hamiltonian (\ref{eq:ham}), 
the dynamics correlates system and environmental degrees of freedom, but this correlation is \emph{classical} at all times $ t > 0 $.

Due to the above classical correlation between system and environment, a trace over the latter induces the decay of the former's coherences, 
on a time scale given by\cite{Breuer2004}
$ t_\text{D} = \left( \sum_k A_k^2 \right)^{-1/2} $, since the different environmental states $\ket{\vb{m}}$ pick up different phases as time evolves -- they {\em dephase}. 
Since these different phases 
nonetheless remain well-defined in the course of time, 
\emph{spin echo} experiments can reverse the effect: Application, at time $ t=t_\text{e} \gg t_\text{D} $, of a $ \pi $ pulse inducing a spin rotation around the $ x $ axis enforces, 
under the same Hamiltonian dynamics, 
a revival of the central spin's coherence at a time $ t = 2t_\text{e} $.

This scenario can be embedded in a more general framework, where 
a system, 
initialized in $ \rho_0 $, evolves 
under a Hamiltonian $ H_\omega $ 
parametrized by a random variable $ \omega $, a proxy, e.g., for disorder 
effects or slow thermal drifts, which changes 
every time 
the experiment is performed. 
If the variations of $ \omega $ are not 
under the experimenter's control (hence ``random''), the system state needs to be described by an
average over the \emph{ensemble} of different realizations $ \rho_\omega(t) =  U_\omega(t) \rho_0 U_\omega^\dagger(t)$, 
with suitably chosen 
probability distribution $ f(\omega) $ \cite{Carnio2015,kropf2016}:
\begin{align}
\bar{\rho}(t) & = \int f(\omega) \rho_\omega (t) \dif \omega \nonumber \\
& = \int f(\omega) U_\omega(t) \rho_0 U_\omega^\dagger(t) \dif \omega \, .
\end{align}
In the particularly simple case where the noise only affects the eigenvalues $ \varepsilon_n^{(\omega)} $ of 
$ H_\omega $, while 
the eigenstates $ \lbrace \ket{\varepsilon_n} \rbrace_n $ 
remain invariant, 
the time-evolved density matrix reads
\begin{equation}
\rho_\omega(t) = \sum_{m,n} r_{m,n} \E^{-\im [\varepsilon_m^{(\omega)} - \varepsilon_n^{(\omega)}]t/\hbar} \ket{\varepsilon_m}\bra{\varepsilon_n} \; ,
\end{equation}
with the $r_{m,n}$ encoding the initial condition $ \rho_0 = \sum_{m,n} r_{m,n} \ket{\varepsilon_m}\bra{\varepsilon_n} $.
The matrix elements of the disorder-averaged state are then 
controlled by the function
\begin{equation}
\gamma_{m,n}(t) = \int f(\omega)  \E^{-\im [\varepsilon_m^{(\omega)} - \varepsilon_n^{(\omega)}]t/\hbar} \dif \omega \; ,
\end{equation}
and it immediately follows that the populations are unaffected by 
the random average,
since $ \gamma_{m,m}(t) \equiv 1, \forall m $. The coherences, instead, 
fade away if
$ \gamma_{m,n}(t) $
vanishes in the limit $ t \rightarrow \infty $. Such asymptotic suppression of the statistical operator's off-diagonal elements, in synchronicity with 
invariant populations, is the formal hallmark of \emph{dephasing}. 
Under the assumption of a long-time behaviour 
$ \gamma_{m,n}(t) \sim \E^{-t/T_2}$, 
$T_2$ is often referred to as dephasing or \emph{transverse relaxation} time \cite{Slichter1996}.

\begin{figure}[t]
\includegraphics[width=\columnwidth]{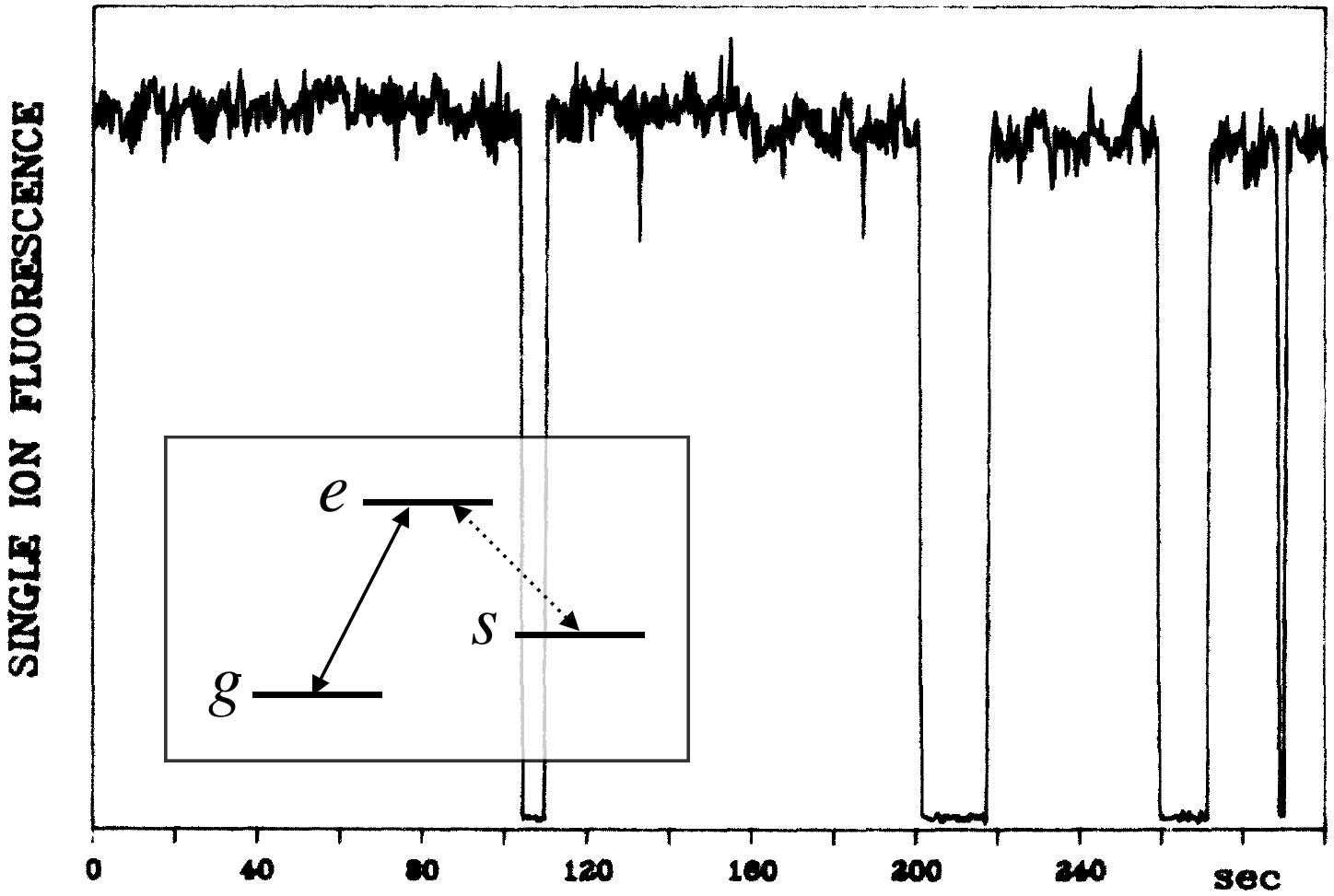}

\caption{
Resonance fluorescence as a function of time for a three-level system depicted in the inset. 
The solid arrow indicates the strong laser-driven transition between the ground state $g$ and 
the excited state $e$, while the dotted arrow indicates the weak transition between e and
the metastable ``shelf'' state $s$. When a $e \rightarrow s$ transition takes
place the fluorescence signal vanishes (see text). Reproduced from Ref.\ \citenum{Sauter1986}.}
\label{fig:quantum-jumps}

\end{figure}

In the above, the coherences of the statistical operator were suppressed by averaging procedures which were formally induced 
by either a trace over the environmental degrees of freedom or over different random realizations of the Hamiltonian system 
dynamics. The density matrix' time evolution thus encodes the full measurement statistics, at any moment in time, of a sufficiently large
ensemble of measurement records. This picture however still suggests that measurement results are accumulated on an ensemble 
of identically and independently prepared quantum systems, while we did insist in our Introduction that quantum mechanical uncertainty is 
an intrinsic property of single quantum objects, in contrast to classical randomness which emerges from the ensemble statistics. 

Only in the 
second half of the 1980's has the latter view been firmly underpinned by the experimental observation \cite{Nagourney1986,Bergquist1986,Sauter1986} of single ions' 
quantum jumps \cite{Schrodinger1952a}.
These experiments involve atomic systems composed of three electronic eigenstates of the bare atom, as shown in Fig.\ \ref{fig:quantum-jumps}. The strong transition between ground and excited states is driven by laser light, but there is also the possibility of a weak transition from the excited to a third metastable state.
While the strong transition is associated with the observation of resonance fluorescence, when a transition into the metastable state takes place the fluorescence radiation vanishes.
Since the precise moment in time when a weak transition into the metastable ``dark'' state occurs is fundamentally unpredictable, the 
detected fluorescence exhibits quantum jumps between random dark and bright periods. Obviously, these jumps would not be observable if
the measurement were performed on an ensemble of three-level atoms, since the individual systems jump at different 
instances, and the effect therefore would be washed out by collecting the total fluorescence. In turn, for the descriptions 
to be consistent, the statistical features of a single atom's time evolution (in jargon: of its ``quantum trajectory'') must be fully 
consistent with the full measurement statistics as encoded by the density matrix, and this is indeed the case.

Quantum jumps are challenging also from the theoretical point of view\cite{meystre1988,Molmer1996,Wiseman2009}, since the state of the atom 
experiences an ``instantaneous'', discontinuous change 
upon registration of a measurement event, because the latter updates our knowledge on the former.
This measurement-induced evolution is eminently 
{\em non-linear and non-unitary,} i.e.\ it is not induced by a Hermitian operator like the Hamiltonian.
If the atom is prepared in an initial state 
$ \ket{\Psi} $, and 
the photodetector subsequently clicks $r$ times, at random moments $\vb \tau = \left( \tau_1, \tau_2, \ldots, \tau_r \right)$,
another state vector $\ket{\Psi_{\vb \tau}}$ 
describes the system state 
after this specific measurement record.
If $f(\vb \tau)$ is the probability of the above specific sequence of events, the state of the system at time $t$ is 
to be described by 
the statistical mixture of all possible evolutions:
$ \rho(t) = \sum_{\vb \tau} f(\vb \tau) \ket{\Psi_{\vb \tau}} \bra{\Psi_{\vb \tau}}$, which, in turn, must coincide with 
the time evolved statistical operator as given, e.g., by (\ref{dynmap}) or by other evolution equations as, e.g., 
the corresponding 
{\em quantum Markovian master equation}.\cite{breuerbook} Because a quantum trajectory as defined above is, by construction,
intimately related to the specific measurement protocol, and since the same system dynamics can be probed
by different measurement set-ups which in general induce different trajectories (or {\em unravellings}), a given 
time evolution of the statistical operator does indeed allow different unravellings.

While our above considerations are valid for quantum systems of arbitrary size and structure, they 
have so far been fully elaborated and validated only in experiments on 
rather small, very well controlled, quantum optical model systems, where disorder can be screened out with high efficiency. When, instead, it comes to 
witnessing quantum uncertainty in \emph{large} molecular structures,
e.g., in photosynthetic light harvesting complexes, or in quantum computing platforms of increasing size,
various forms of disorder are at play
and experiments on single realizations of such structures do only since recently allow to isolate 
distinct sources of dephasing-induced decoherence \cite{Hildner2013}. 

Given what is known from 
quantum transport in disordered systems \cite{akkermans2011} and from quantum chaos \cite{bohigas1991,gutzwiller1991}, it is 
clear that \emph{static} disorder, i.e.\ variations in the complex structural properties of these large molecular functional units,
will imprint large, interference-induced (and therefore deterministic) 
fluctuations on typical observables when sampling over different realizations \cite{walschaers2013}.
Furthermore, since 
biological macro-molecular structures 
typically function at physiological temperatures and are an interconnected part of living matter, they are subject to fluctuations on time scales usually shorter than those of interest (\emph{dynamical} noise). Any non-trivial quantum dynamical feature of such systems, which are effectively far from equilibrium, must be averaged over these fast fluctuations and can thus only be transient, namely it {\em must} fade away on finite time scales, if not compensated for by some pumping mechanism.\cite{shatokhin2018}

If we consider that functionality manifests on time, length and energy scales much larger than those defined by static and dynamical disorder,
the quest for quantum-enhanced functionality therefore seems a highly non-trivial endeavor with open outcome.
However, while much of the known 
phenomenology of quantum dynamics in macromolecular complexes at ambient temperatures may suggest that quantum phenomena are unlikely to determine functionality,\cite{miller2017}
it appears worthwhile to consider examples from other areas: from coherent backscattering of light, first observed (without being understood), 
more than 100 years ago, on Saturn's rings \cite{mueller1893}, over
random lasing \cite{Letokhov1968,Wiersma2008}, a paradigmatic quantum coherence phenomenon in {\em widely open} systems driven far from equilibrium, to the directed transmission of quantum states across
disordered scattering media\cite{gigan2016,Leonhard2018,wellens2012} and, of course, Anderson localization-induced metal-insulator transitions \cite{Anderson1958a,Kramer1993,Evers2008}. While most of these examples do not share the intricate 
interplay of a hierarchy of length, energy and time scales as one of the most intriguing features of living matter, they do offer convincing evidence against the 
slightly too reductionist point of view that quantum coherence can only persist and manifest under highly controlled conditions.

What qualitatively changes with respect to those simple systems on which most of the existing theory of quantum coherence and decoherence was tested is the 
ubiquity of the above-mentioned fluctuations, due to the interplay of strongly coupled degrees of freedom, and the high spectral densities thereby generated. The (eigen-)mode structure of such molecular 
systems 
is typically highly sensitive to small changes of the boundary conditions (which change from single realization to single realization), with this sensitivity itself 
being a manifestation of quantum interference effects.

It therefore appears imperative to adopt a statistical point of view, where the statistics 
of measurement records from different individual samples is explored as an additional source of information, as well as a potential 
resource.\cite{Wiersma2008,walschaers2013,Hildner2013,krueger2016} This 
even more so 
because biological matter typically displays a high degree of redundancy on the microscopic level, such that functionality on large scales appears likely to be determined by cooperative 
effects which rely on that very redundancy.
Naturally follows the question 
under which conditions 
certain functionally relevant properties of particular eigenmodes (e.g.\ identified by their overlap with a 
pre-selected initial or target state, or by a typical injection energy) 
are statistically sufficiently abundant, such that sampling over a
sufficient number of realizations can guarantee a certain level of (robust) functionality.\footnote{Note from the at least at a first glance rather distinct scenario of quantum annealing 
platforms that some of the successful applications of these
do precisely rely on such statistical considerations.\cite{king2018}}
Such statistical considerations need to be underpinned by an assessment of the competition of the desired coherent processes with environment-induced decoherence and loss 
mechanisms, as well as with external driving forces.

Recall from our above elementary discussion of quantum jumps that wave-particle duality can very well be fully expressed on the level of single realizations while completely obliterated by ensemble averages. What is therefore needed are theoretical and experimental tools which allow to induce and decipher single realizations of quantum dynamical processes in isolated molecular samples with statistically characterized structural properties.
On the theory side, this will require a generalization of standard quantum optical and open systems methods for many degrees of freedom systems (where the many degrees 
of freedom can be those of a single as well as of a many-particle configuration space) with an increased number of coupling agents and decay channels.
On the experimental side, probing the subtle quantum features of complex systems requires correspondingly subtle probing schemes, such as those offered by quantum optical tools. This, in particular, implies the need for quantum probes, as offered, most prominently and with enormous versatility, by quantum states of light tailored in {\em all} degrees 
of freedom.\cite{raymer2013,dorfman2016}

\subsection{Acknowledgment}
Supported by the Georg H. Endress foundation.

\providecommand{\latin}[1]{#1}
\makeatletter
\providecommand{\doi}
  {\begingroup\let\do\@makeother\dospecials
  \catcode`\{=1 \catcode`\}=2 \doi@aux}
\providecommand{\doi@aux}[1]{\endgroup\texttt{#1}}
\makeatother
\providecommand*\mcitethebibliography{\thebibliography}
\csname @ifundefined\endcsname{endmcitethebibliography}
  {\let\endmcitethebibliography\endthebibliography}{}

\end{document}